# Templated Dewetting-Alloying of NiCu Bilayers on TiO$_2$ Nanotubes Enables Efficient Noble Metal-Free Photocatalytic H$_2$ Evolution

by Davide Spanu,[a,b] Sandro Recchia,[b] Shiva Mohajernia,[a] Ondřej Tomanec,[c] Štěpán Kment,[c] Radek Zboril,[c] Patrik Schmuki[a,c,d]* and Marco Altomare[a]*

[a] Department of Materials Science and Engineering WW4-LKO, University of Erlangen-Nuremberg, Martensstrasse 7, D-91058 Erlangen, Germany

[b] Department of Science and High Technology, University of Insubria, via Valleggio 11, 22100 Como, Italy

[c] Regional Centre of Advanced Technologies and Materials, Faculty of Science, Palacky University, Olomouc, Šlechtitelů 27, 783 71 Olomouc, Czech Republic

[d] Chemistry Department, Faculty of Sciences, King Abdulaziz University, 80203 Jeddah, Saudi Arabia Kingdom

* Corresponding author. Email:     marco.altomare@fau.de

schmuki@ww.uni-erlangen.de

Link to the published article:

https://pubs.acs.org/doi/abs/10.1021/acscatal.8b01190

**Abstract**

Photocatalytic $H_2$ evolution reactions on pristine $TiO_2$ is characterized by low efficiencies due to trapping and recombination of charge carriers, and due to a sluggish kinetics of electron transfer. Noble metal (mainly Pt, Pd, Au) nanoparticles are typically decorated as cocatalyst on the $TiO_2$ surface to reach reasonable photocatalytic yields. However, owing to the high cost of noble metals, alternative metal cocatalysts are being developed. Here we introduce an approach to fabricate an efficient noble metal free photocatalytic platform for $H_2$ evolution based on alloyed NiCu cocatalytic nanoparticles at the surface of anodic $TiO_2$ nanotube arrays. NiCu bilayers are deposited onto the $TiO_2$ nanotubes by plasma sputtering. A subsequent thermal treatment is carried out that leads to dewetting, that is, owing to surface diffusion the Ni and Cu sputtered layers simultaneously mix with each other and split into NiCu nanoparticles at the nanotube surface. The approach allows for a full control over key features of the alloyed nanoparticles such as their composition, work function and cocatalytic ability towards $H_2$ generation. Dewetted-alloyed cocatalytic nanoparticles composed of equal Ni and Cu amounts not only are significantly more reactive than pure Ni or Cu nanoparticles, but also lead to $H_2$ generation rates that can be comparable to those obtained by conventional noble metal (Pt) decoration of $TiO_2$ nanotube arrays.

**Introduction**

Since the ground-breaking work of Fujishima and Honda[1] in 1972, the photocatalytic production of $H_2$ by splitting of $H_2O$ has been widely investigated. Among the different photocatalytic materials developed in the last decades, $TiO_2$ still represents one of the most explored and promising semiconductor metal oxides. In fact, not only is $TiO_2$ cheap, easily available, non-toxic, and stable against corrosion and photo-corrosion,[2,3] but most importantly, it has a suitable conduction band edge position to allow $H_2O$ reduction to $H_2$, the fuel of the future. For $TiO_2$, the conduction band (CB) edge lies 0.45 eV higher than the redox potential of water.[4] Therefore, photo-generated conduction band electrons, promoted by UV light irradiation ($E_{g\ TiO_2} \sim$ 3.0-3.2 eV), are thermodynamically able (in terms of electron "exit" energy) to reduce $H_2O$ to $H_2$. However, this reaction is characterized by low efficiencies owing to trapping and recombination of charge carriers, and due to a sluggish kinetics of electron transfer from $TiO_2$ to reactants.

A successful strategy to limit charge recombination in $TiO_2$ is provided by nanostructuring the semiconductor. Particularly, one-dimensional (1D) nanostructures, such as anodic $TiO_2$ nanotubes (NTs), have attracted great attention in the last years.[5,6] Aligned arrays of self-organized NTs with controllable morphology can be grown by a simple electrochemical anodization of Ti metal in a suitable electrolyte.[2,3,7] These highly-ordered and defined 1D structures promote directional charge transport and orthogonal electron–hole separation that lead to enhanced photocatalytic and photo-electrochemical efficiencies, owing to a more efficient charge separation and improved electron transport properties.[8]

On the other hand, limitations in charge transfer can be tackled by using a suitable cocatalyst, mostly noble metal nanoparticles (NPs), e.g. Pt, Au and Pd.[9] The most efficient metal-cocatalyst is Pt, which not only enables an efficient electron transfer at the $TiO_2$/environment interface yielding a favorable solid state (Schottky type) junction to $TiO_2$,[4] but also provides catalytic sites that enhance the hydrogen recombination reaction ($2H^0 \rightarrow H_2$).[10]

However, the high cost of Pt (or in general of noble metals) has meanwhile shifted the attention to alternative catalysts and cocatalysts composed of non-noble and cheaper metals. Interesting candidates are e.g. Ni and Cu. Nevertheless, the photocatalytic activity of $TiO_2$ decorated with non-noble metal cocatalysts is typically lower than that achieved by Pt decoration (under comparable deposition conditions).[11–15]

In recent work, photocatalysts composed by co-loaded Ni and Cu NPs on powdered titania were shown to enhance the $H_2$ evolution rate more than pure Ni or Cu particles.[16–19] Coloaded NiCu NPs found also promising application in sensors,[20,21] dye photo-degradation processes[22] and selectively catalyzed organic reactions.[23–25] NiCu-$TiO_2$ structures are commonly prepared by a powder technology approach based on wet impregnation,[22,24] hydrothermal deposition,[17,25] thermal deposition followed by a chemical reduction,[18,23] laser ablation in liquid[16] and electrodeposition.[20]

Here we propose an alternative approach based on a templated dewetting-alloying principle of sputtered Ni and Cu metal bilayers on a highly defined $TiO_2$ nanotube surface.[26,27] The amount of Ni and Cu metal can be adjusted by the thickness of the metal sputtered layers (nm range). With a suitable thermal treatment, the metal films can, owing to surface diffusion, mix (alloy) with each other and simultaneously dewet, i.e. split into NiCu NPs of controllable composition, size and spacing,[27] at the surface of $TiO_2$ nanotubes. We show that dewetted-alloyed NiCu nanoparticles of an optimized composition lead to a significantly higher photocatalytic $H_2$ evolution efficiency compared to pure Ni or Cu nanoparticles decorated on the nanotubes under similar conditions. Even more remarkable, the $H_2$ generation rate of NTs decorated with dewetted-alloyed NiCu NPs is comparable with that obtained by Pt decoration (under optimized Pt deposition conditions), particularly under solar light irradiation.

**Results and Discussion**

In the present work we use short aspect ratio $TiO_2$ NTs as photocatalytically active surface and patterned substrate for dewetting. The $TiO_2$ NT arrays are grown on Ti foils by self-organizing electrochemical anodization in a hot $H_3PO_4$/HF electrolyte.[26] Figure 1a and Figure S1a shows that these NTs are almost ideally hexagonally packed, and each NT has an individual inner diameter of ~ 90 nm, a depth of ~ 180 nm and wall thickness of ~ 10 nm. These short and well-defined nanotubes can easily be coated with metal films (Ni and Cu in this work) by a plasma sputtering technique.[28,29]

Examples of the resulting structures are illustrated in Figure 1b and Figure S1b,c. These structures are formed by sputter-coating on the $TiO_2$ NTs a Cu film (nominal thickness 10 ± 0.5 nm), a Ni film (10 ± 0.5 nm) or a NiCu bilayer deposited by sequential sputtering (5 ± 0.5 nm for each metal, for an overall nominal thickness of ∼ 10 nm). In any case the as-sputtered metal films homogeneously coat the $TiO_2$ NT surface.

Afterwards, a thermal treatment at 450°C (1 h) leads to dewetting (Figure 1c-e) of the metal films at the $TiO_2$ NT surface. The optimization of the annealing/dewetting parameters is based on the results of previous work – more details are in the SI. Dewetting takes place as thin metal films are unstable when heated up, and tend to split-open and agglomerate forming metal islands (particles) via surface diffusion. Dewetting of a given metal thin film typically initiates at temperatures between 1/3 and 2/3 of the metal melting point[30]; thus, the occurrence of Ni and Cu dewetting at 450°C is well in line with their melting point ($T_m$ Cu = 1085°C; $T_m$ Ni = 1455°C). Such thermal treatment leads at the same time to the crystallization of the NT substrate into a mixed anatase-rutile $TiO_2$ phase (see XRD data in Figure S2).

The average size of the dewetted metal particle is found to be independent of the initial composition of the metal film. Both Ni and Cu layers, as well as NiCu bilayers, dewet into ~ 30-40 nm-sized NPs (see NP size distribution in Figure S3a-c). Compared to the dewetted Ni NPs that are relatively close to each other and homogeneously decorated on the substrate, the dewetted Cu NPs are less densely distributed and more scattered at the $TiO_2$ NT surface. This may suggest that Ni and Cu exhibit slightly different dewetting "modes", e.g. owing to a different surface diffusion or substrate wettability, and the dewetting behavior of NiCu bilayers may depend on the initial composition of the bilayer.

Key role in tailoring (i.e. limiting) the size of the dewetted metal particles to a few tens of nm is played by the patterned $TiO_2$ NT surface that provides high surface energy sites, such as the NT rims, that act as preferential locations that initiate dewetting – in other words, the defined $TiO_2$ NT morphology enables templated-dewetting.[30] Note in fact that, on a flat $TiO_2$ surface, NiCu bilayers of comparable composition and thickness split into particles that can be as large as ca. 300 nm, with an average size > 100 nm that is one order of magnitude bigger than on the $TiO_2$ NT surface (see Figure S3d,e). Such results are obviously of large interest e.g. in the field of heterogeneous catalysis, where metal particle size, distribution and composition are crucial parameters for reactivity control.[25]

By systematically varying the Ni and Cu relative amount (i.e. nominal thickness), we prepared a series of $TiO_2$ NT arrays decorated with dewetted NiCu particles with defined compositions – a summary of the prepared samples and nominal cocatalyst composition is in Table S1, while their morphology is illustrated in Figure S4. The structures were investigated as photocatalyst for $H_2$ evolution from ethanol-water under UV light illumination (365 nm, LED, 105 mW cm$^{-2}$).

The photocatalytic results are summarized in Figure 2a. The data show that pristine $TiO_2$ NTs ("$TiO_2$") exhibit a negligible photocatalytic $H_2$ generation (0.04 µL h$^{-1}$ cm$^{-2}$). $TiO_2$ NTs modified with dewetted pure Cu or Ni NPs ("10Cu" and "10Ni") show a slightly improved $H_2$ generation rate (1.62 µL h$^{-1}$ cm$^{-2}$ and 0.44 µL h$^{-1}$ cm$^{-2}$, respectively).

Remarkably, cocatalytic NPs formed by dewetting NiCu bilayers (7Ni3Cu, 5Ni5Cu and 3Ni7Cu) lead in any case to a clearly higher photocatalytic efficiency compared to pure Ni or Cu NPs (using a similar loading). The improvement in $H_2$ generation correlates with the Ni-Cu relative amount, and is maximized when Ni and Cu are in equal amounts; the highest photocatalytic activity is measured for $TiO_2$ NT structures decorated by dewetting a sequentially sputtered 5 nm Ni – 5 nm Cu bilayer ("5Ni5Cu-$TiO_2$"). These structures lead to a $H_2$ evolution rate of 6.04 µL h$^{-1}$ cm$^{-2}$, which is ~ 14 and ~ 4 times higher than that of "10Ni" and "10Cu", respectively. We also found that in this case the sputter deposition sequence of Ni and Cu metal films does not play a crucial role. Sample 5Cu5Ni was prepared by using an opposite metal sputter-deposition sequence compared to sample 5Ni5Cu. Both sample lead however to almost identical $H_2$ evolution rates, i.e. 5.60 and 6.04 µL h$^{-1}$ cm$^{-2}$, respectively. The apparent quantum efficiency (AQE) for the most active photocatalyst (e.g. 5Ni5Cu) under monochromatic UV light illumination is ~ 0.043% (additional details are in the SI).

The results of a 30 h long photocatalytic run (Figure 2b) reveal that $H_2$ is evolved at a constant rate. Also, SEM images of the photocatalyst taken after 30 h long photocatalytic runs are virtually identical to those of the as-prepared (not used) photocatalyst (Figure S5). This proves the high stability of the NiCu-TiO$_2$ photocatalysts, and deterioration phenomena such as (photo-)corrosion or fall-off decay of the metal cocatalyst could not be observed.

The data in Figure 2c show that the photocatalytic improvement is a direct consequence of dewetting. The morphology of structures treated in Ar at 400°C is virtually identical to that of as-sputtered materials, i.e. such temperatures do not induce dewetting of the NiCu bilayer, and the resulting H$_2$ generation rate is significantly lower than that of photocatalysts dewetted at 450°C. Some key reasons for the higher activity obtained upon annealing-dewetting at 450°C are: (i) dewetting exposes free TiO$_2$ surface that is crucial for the hole-mediated reaction (ethanol oxidation); (ii) metal nanoparticles formed by dewetting are alloyed while the as-sputtered film is a Ni-Cu bilayer (discussed below); (iii) dewetted metal NPs exhibit a higher surface area compared with as-sputtered metal films; (iv) dewetting forms localized metal/oxide Schottky junctions at the TiO$_2$ NT surface; and (v) upon dewetting, shading effects of the metal cocatalyst are reduced and a higher photon flux can reach the TiO$_2$ substrate. Nevertheless, although the annealing at 500°C also induces dewetting, it leads to a relatively low H$_2$ evolution performance – as reported by Yoo et al.,[31] this is due to an extensive formation of TiO$_2$ rutile phase in the NTs for annealing T > 450°C.

Reference samples fabricated by dewetting smaller amounts of Cu or Ni, e.g. samples "5Cu" and "5Ni", lead to H$_2$ generation rates (0.41 µL h$^{-1}$ cm$^{-2}$ and 0.19 µL h$^{-1}$ cm$^{-2}$, respectively) that are even lower than that of 10Cu and 10Ni (Figure 2a). This means that the improved activity of TiO$_2$ NTs decorated with dewetted NiCu NPs cannot be simply ascribed to the Ni or Cu loading. Moreover, one may consider that if Ni and Cu could ideally be deposited on the TiO$_2$ NTs without undergoing alloying, the resulting H$_2$ evolution rate would virtually be the sum of those of samples 5Ni and 5Cu, i.e. 2.1 µL h$^{-1}$ cm$^{-2}$ which is still ~ 3 times lower than that obtained by alloying similar

Ni and Cu cocatalyst amounts (sample 5Ni5Cu leads to ~ 6 µL h$^{-1}$ cm$^{-2}$) – this results support the synergy enabled by alloying Ni and Cu to form NiCu alloyed nanoparticles as TiO$_2$ cocatalyst for noble metal free photocatalytic H$_2$ evolution.

To clarify the reason for the photocatalytic improvement, we characterized the different samples by UV-Vis diffuse reflectance spectra, XRD, XPS and TEM, particularly in view of the morphology and physicochemical features of the dewetted Ni, Cu of NiCu particles.

The UV-Vis diffuse reflectance spectra of the different Ni-, Cu- and NiCu-decorated TiO$_2$ nanotubes are shown in Figure S6. While the spectra differ from each other, probably as a consequence of the different composition of the metal cocatalyst nanoparticles, the trend of the light absorption at 365 nm (i.e. λ used for the photocatalytic experiments) does not support the trend of photocatalytic efficiency (see Figure S6b) – in other words, the photocatalytic enhancement obtained by TiO$_2$ decoration with alloyed NiCu nanoparticles cannot be dominated by light absorption features.

XRD results are compiled in Figure 2d. For TiO$_2$ NTs decorated with dewetted Cu NPs (10Cu), the peaks at 43.3° and 50.4° are indexed as the Cu (111) and (200) planes.[32] For Ni NP decorated TiO$_2$ NT (10Ni), the XRD signals at 44.4° and 51.9° are assigned to Ni (111) and (200) diffraction peaks.[33]

Interestingly, for TiO$_2$ NTs decorated with dewetted NiCu particles of various compositions, the (111) diffraction peak shifts from 43.3° (pure Cu) to 44.4° (pure Ni) with increasing the Ni loading, while no reflection of pure Ni or Cu could be detected. In other words, the (111) diffraction signal for dewetted NiCu particles shows a shift that correlates well with the initial composition of the NiCu bilayer. These results fit well the Vegard's law[16,34] that describes the correlation between the XRD peak position of a binary AB metal alloy with respect to the content of the alloying metal elements A and B (see Figure S7 and data in Table S2). Thus, one can conclude that dewetting of NiCu bilayers at the surface of $TiO_2$ NTs forms NiCu alloyed NPs. These results are in principle well in line also with the full miscibility for any composition of the NiCu alloy system. The results of the fitting based on Vegard's law (Figure S7 and Table S2) suggest that the composition of the alloyed NPs is in line with that of the NiCu bilayer – the small deviation from the predicted trend (Inset in Figure 2d and Figure S7) may either originate from the fact that the thermodynamics of bulk NiCu alloys may not apply in the same way to nanosized alloyed particles or could be ascribed to the observed slight compositional inhomogeneity.

The improved photocatalytic activity of $TiO_2$ NTs decorated with dewetted-alloyed NiCu NPs is therefore attributed on the one hand to the specific cocatalyst composition that can lead to a faster H atom adsorption/recombination over Ni cocatalytic centers combined with a more favorable $H_2$ molecule desorption from adjacent Cu sites[35,36] – see Scheme S1. This mechanism can also take place in the case of NiCu particles with an inhomogeneous composition: H atoms may adsorb and recombine efficiently at Ni rich sites, while desorption of $H_2$ molecules may occur from the nearby Cu rich sites in the same cocatalyst nanoparticle.

On the other hand, an additional reason for the improved activity is the work function of NiCu alloyed particles with respect to that of pure Cu and Ni (i.e. 4.65 eV and 5.15 eV[37], respectively). According to theoretical calculations,[38–40] the work function of NiCu alloy increases linearly from 4.65 eV to 5.15 eV with increasing the Ni content. This leads in principle to an increase of the Schottky barrier height at the $NiCu/TiO_2$ interface, which can enable a more efficient electron-hole separation – see Scheme S2. As the highest rate of $H_2$ evolution is achieved with a 1:1 Ni:Cu ratio, one can conclude that such composition of the NPs enables at the same time an efficient electron-hole separation (due to a higher Schottky barrier height compared to pure Cu), and a fast kinetics of hydrogen adsorption/recombination/desorption at the NiCu alloy surface.

We also performed $H_2$ evolution experiments with the most active photocatalyst under illumination with unfiltered and filtered (420 nm cut-off filter) simulated solar light (AM1.5G). As shown in Figure 2e, under filtered illumination ($\lambda > 420$ nm) the NiCu-TiO$_2$ photocatalyst exhibits a negligible $H_2$ generation (0.02 µL h$^{-1}$ cm$^{-2}$), while the activity is significantly higher (0.28 µL h$^{-1}$ cm$^{-2}$) under full lamp illumination. This indicates that the photocatalyst is active only under UV light illumination ($\lambda < 420$ nm). Thus, the light absorber and charge carrier generator is TiO$_2$, while the NiCu NPs act as electron transfer mediator and hydrogen evolution site.

A comparison between the photocatalytic activity of NiCu-TiO$_2$ NTs with noble metal decorated TiO$_2$ structures is shown in Figure 2f,g. As benchmark photocatalyst we fabricated Pt decorated TiO$_2$ NTs. The approach used for the fabrication of Pt-TiO$_2$ structures is in principle the same used for NiCu-TiO$_2$, i.e. a Pt film of an optimized thickness (1 nm, based on previous work)[41] is sputter-coated on TiO$_2$ nanotube arrays that are then annealed at 450°C in argon – this treatment simultaneously leads to nanotube crystallization and Pt dewetting into nanoparticles. The $H_2$ evolution rate of NiCu-TiO$_2$ NTs is ~ 60% and 50% of that measured for Pt-TiO$_2$ NTs under solar light illumination with and without the 420 nm cut off filter, respectively. This result is remarkable if one takes into account the larger abundance of Ni and Cu and the substantially lower fabrication cost of the NiCu cocatalyst.

The most active photocatalyst (5Ni5Cu-TiO$_2$) was further investigated by HAADF-TEM and EDS-TEM. The HAADF-TEM images in Figure 3a show that the dewetted metal particles at the TiO$_2$ NT surface are ~ 30-40 nm in size. This result matches well the size distribution estimated from the SEM images (Figure 1e and Figure S3). From the EDS-TEM mapping data in Figure 3b, it seems that each particle results from welding of two different entities, one mainly composed of Cu, and the other of Ni. However, the zone where the two entities are welded into a single particle shows a mixed NiCu composition – see Figure 3c-e. This is well in line with the XRD results supporting the formation of NiCu alloyed NPs.[42] Please also note that in spite of the partially inhomogeneous composition of these particles, no XRD peak attributable to pure Ni or pure Cu phase could be detected (discussed above).

The EDS-TEM mapping in Figure 3e, suggests that the NiCu metal particles may be partially oxidized at their surface, probably due to surface oxidation of the metal NPs under ambient conditions or due to interaction with the $TiO_2$ substrate.[25] These structures were further characterized by XPS and compared to non-alloyed Ni and Cu NPs on $TiO_2$ NTs. The results are compiled in Figure 3f,g, where the Ni2p and Cu2p signals are deconvoluted into various contributions. While the signals of Ni and Cu oxides and hydroxide are present in any structure, the bands at 852.1 and 931.9 eV in the spectra of sample 5Ni5Cu can be assigned to Ni and Cu metals, respectively.[25,43–45] Both the Ni and Cu metal signals for the dewetted-alloyed NiCu NPs show a positive shift of 0.2 eV, compared to particle dewetted from pure Ni and Cu layers. These results, along with XRD and TEM data, further corroborate the formation of NiCu alloyed NPs.

**Conclusions**

We introduced an approach to fabricate an efficient noble metal free photocatalytic platform for $H_2$ evolution formed by dewetting NiCu bilayers into alloyed NiCu cocatalytic nanoparticles at the surface of $TiO_2$ nanotube arrays. This strategy allows for a full control over the alloyed nanoparticle size, loading and distribution, which enables the tuning of the metal cocatalyst work function as well as its catalytic ability in terms of kinetics of hydrogen adsorption, recombination and desorption. We showed that an equal Ni and Cu content in the alloyed nanoparticles is key to achieve a strong enhancement of the photocatalytic $H_2$ evolution compared to $TiO_2$ decorated with pure Ni or Cu nanoparticles. The alloyed NiCu cocatalyst on $TiO_2$ nanotubes allows to reach $H_2$ generation rates that are comparable to those delivered by conventional noble metal (Pt) decoration of $TiO_2$. From a more general point of view, it is anticipated that the dewetting-alloying approach can be adapted to other alloy systems and will therefore find wide application e.g. in heterogeneous catalysis.

**Acknowledgments**

The authors would like to acknowledge ERC, DFG and the DFG cluster of excellence EAM for financial support. Davide Spanu and Sandro Recchia gratefully acknowledge financial support from MIUR. The authors gratefully acknowledge the support by the Operational Programme Research, Development and Education – European Regional Development Fund, project no. CZ.02.1.01/0.0/0.0/15_003/0000416 of the Ministry of Education, Youth and Sports of the Czech Republic. Helga Hildebrand and Ulrike Marten-Jahns are acknowledged for valuable technical help.

**Supporting Information**

The experimental section and additional catalyst characterization data (SEM, XRD) are provided in Figures S1−S7, Scheme S1-S2 and Tables S1−S2.

**Table of Contents**

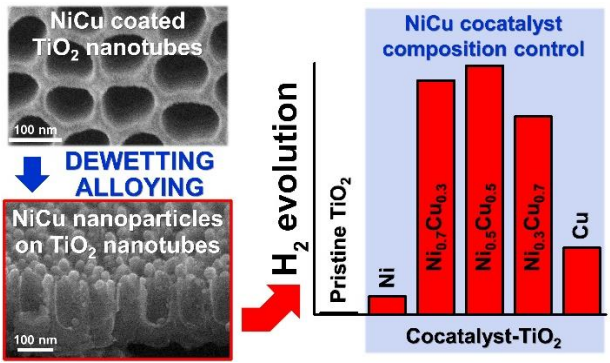

**Figure Captions**

**Figure 1**

SEM images of various $TiO_2$ nanotube arrays. (a) pristine (Inset: cross-sectional view); (b) coated with a NiCu bilayer (5 nm Ni – 5 nm Cu); (c) coated with a 10 nm thick Ni layer and dewetted at 450°C; (d) coated with a 10 nm thick Cu layer and dewetted at 450°C; (e,f) coated with a NiCu bilayer (5 nm Ni – 5 nm Cu) and dewetted at 450°C (f shows the cross-sectional view).

**Figure 2**

(a) photocatalytic $H_2$ evolution rate of different photocatalysts under UV light illumination (LED, 365 nm) as a function of loading, composition and sputter-deposition sequence of the metal cocatalyst; (b) $H_2$ amount evolved over time under UV light illumination (LED, 365 nm) for sample 5Ni5Cu dewetted at 450°C; (c) photocatalytic $H_2$ evolution rate for sample 5Ni5Cu under UV light illumination (LED, 365 nm) as a function of the annealing temperature; (d) XRD patterns of $TiO_2$ nanotube decorated with Ni, Cu or alloyed NiCu NPs of various compositions. Inset: shift of the (111) XRD reflection as a function of the composition of the NiCu alloy NPs; (e) photocatalytic $H_2$ evolution rate for sample 5Ni5Cu (dewetted at 450°C) under unfiltered and filtered (420 nm cut-off filter) simulated solar light illumination (AM1.5G); (f,g) photocatalytic $H_2$ evolution rate for sample 5Ni5Cu (dewetted at 450°C) compared to a Pt- $TiO_2$ benchmark photocatalyst under (f) UV light (LED, 365 nm) and (g) simulated solar light illumination (AM1.5G).

**Figure 3**

(a-e) TEM data for sample 5Ni5Cu dewetted at 450°C: (a,d) HAADF-TEM images; (b,c,e) EDS-TEM images with (b,c) Cu and Ni and (e) O elemental distribution. (f,g) XPS spectra for sample 5Ni5Cu, 10Ni and 10Cu (dewetted at 450°C) in the (f) Ni2p and (g) Cu2p region. The spectra are deconvoluted and peaks of Ni and Cu metal, oxide and hydroxide are indicated. The Cu2p3/2 signal at 931.7-931.9 eV (g) is attributed to Cu metal (based on XRD and TEM results) although the Cu2p3/2 signal of Cu(I) oxide typically peaks at similar B.E. values.

**Figures**

**Figure 1**

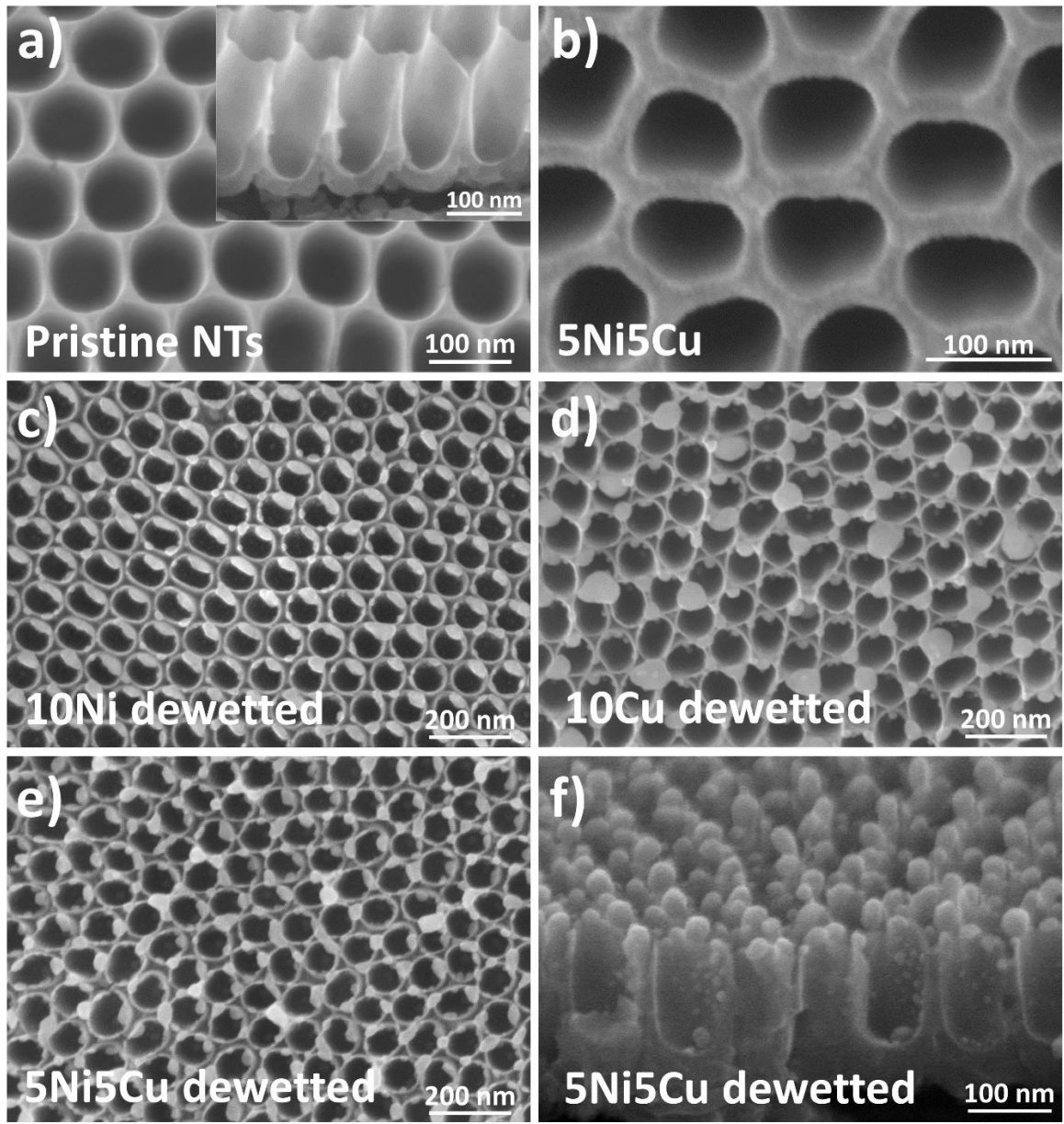

**Figure 2**

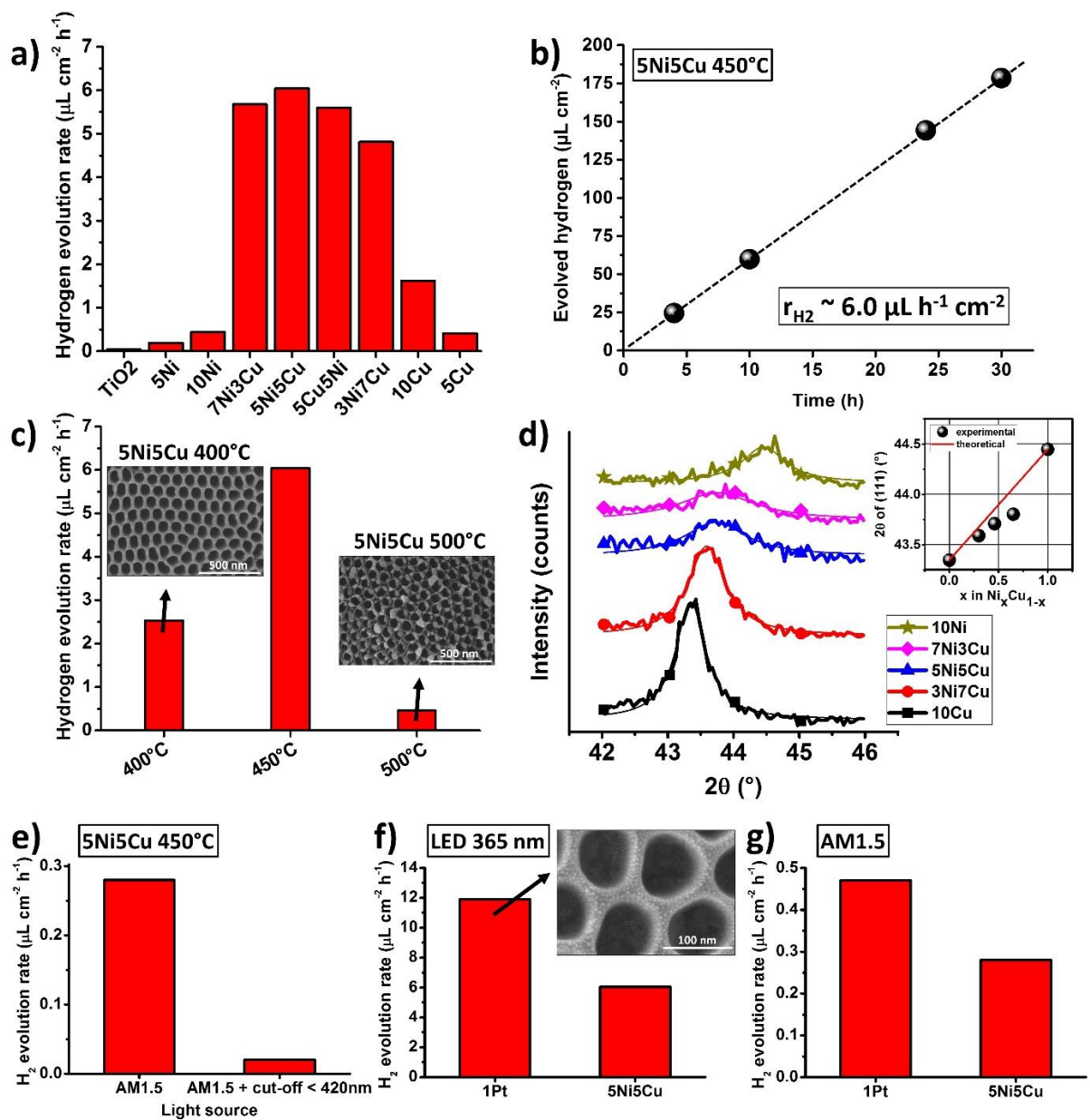
23

**Figure 3**

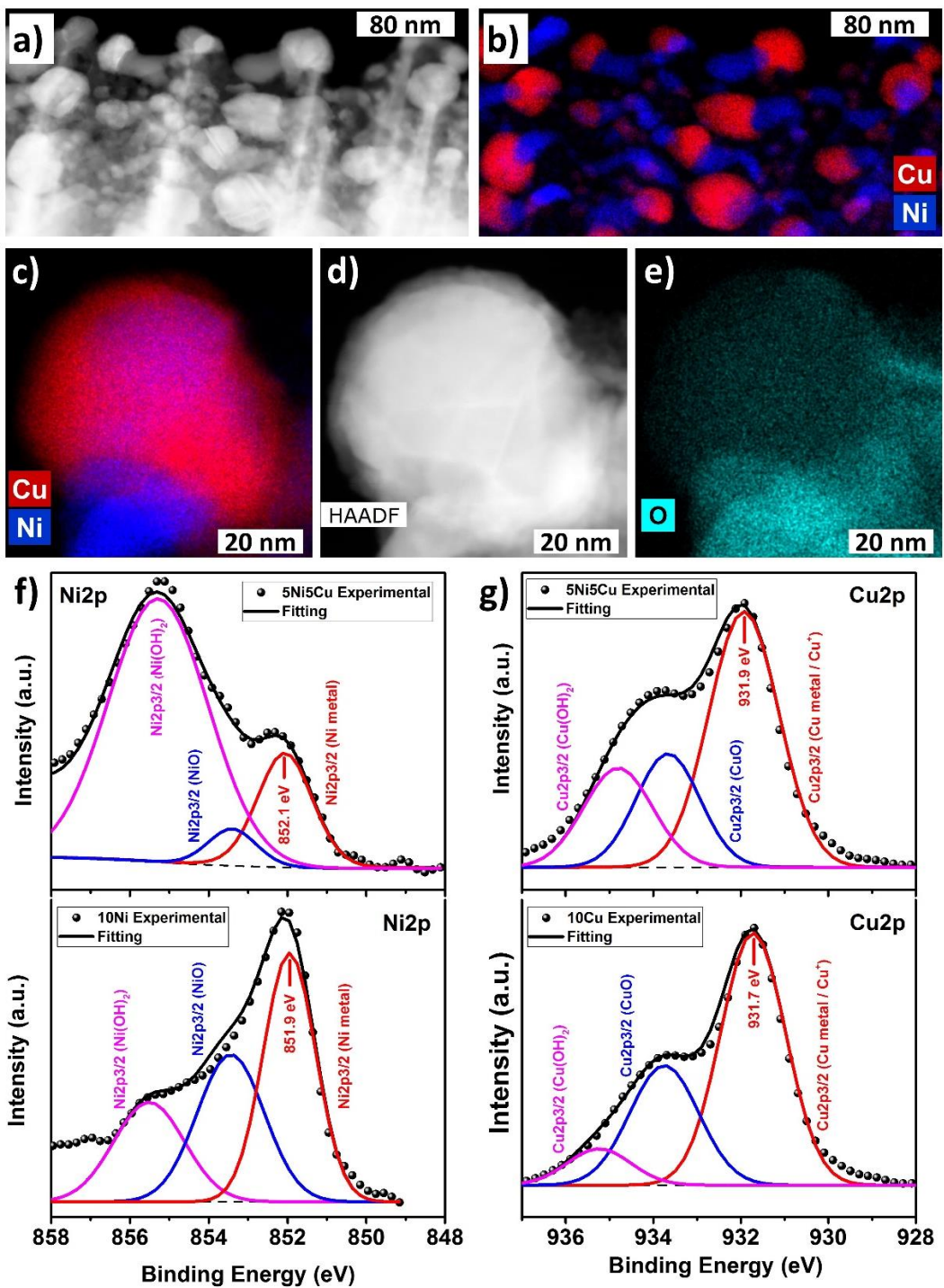